\def\ggf{\leavevmode\hbox{\raise.4ex\hbox{%
    $\scriptscriptstyle<\!<\;$}}}
\def\gdf{\leavevmode\hbox{\raise.4ex\hbox{%
    $\scriptscriptstyle\;>\!>\;$}}}
\def\leftnote#1{\leavevmode\vadjust{\setbox1=\vtop{\hsize 20mm
    \parindent=0pt\small\baselineskip=9pt
    \rightskip=4mm plus 4mm#1}
    \hbox{\kern-2cm\smash{\box1}}}}

\def\inst#1{\unskip$^{#1}$}
\def\email#1{{\tt#1}}

\documentclass[twoside]{article}
\usepackage[T1]{fontenc}
\usepackage[latin1]{inputenc}
\usepackage[francais]{babel}
\usepackage{aeguill}
\usepackage{perpi2006}

\usepackage{epsfig,shadow}
\usepackage{graphicx}
\usepackage{amsmath}
\usepackage{amssymb}

\begin{document}

\parindent=0pt

\title{\Large\bf Étude de performance des systèmes de découverte de ressources}
\shorttitle{}
\author{Heithem Abbes\inst{1,2}\hspace{2em} Christophe Cérin\inst{2}\hspace{2em}
Jean-Christophe Dubacq\inst{2} \hspace{2em} Mohamed Jemni\inst{1}}

\address{\inst{1}\'Ecole supérieure des sciences et techniques de Tunis, Unité de recherche UTIC \\
 5, Av. Taha Hussein, B.P. 56, Bab Mnara, Tunis, Tunisia \\
 Tel: (+216) 71 496 066    Fax: (+216) 71 391 166 \\
\email{heithem.abbes@esstt.rnu.tn} \hspace{2em}
\email{mohamed.jemni@fst.rnu.tn}}
\address{\inst{2}LIPN, UMR 7030, CNRS, Université Paris-Nord\\
99, avenue Jean-Baptiste Clément, 93430 Villetaneuse, France\\
Tel: +33-(0)1.49.40.35.78   Fax : +33-(0)1.48.26.07.12
\email{\{christophe.cerin,jean-christophe.dubacq\}@lipn.univ-paris13.fr}}
\bibliographystyle{plain}
\maketitle

\Resume{Les grilles de PC (Desktop Grid) sont une technologie qui
  consiste à exploiter des ressources géographiquement dispersées, pour
  traiter des applications complexes demandant une grande puissance de
  calcul et une capacité de stockage importante.  Cependant, comme le
  nombre de ressources augmente, les besoins de changement d'échelle,
  d'auto-organisation, de reconfiguration dynamique, de décentralisation
  et de performance deviennent de plus en plus indispensables. Comme ces
  propriétés sont présentes dans les systèmes P2P (pair-à-pair), la
  convergence des grilles et des systèmes P2P semble naturelle. Dans ce
  contexte, l'article évalue l'adaptation au changement d'échelle et la
  performance des outils P2P pour la publication/découverte de
  services. Trois bibliothèques sont évaluées à cet effet: Bonjour,
  Avahi et Pastry. Nous étudions leur comportement vis à vis des
  critères qui sont le temps écoulé pour l'enregistrement des services
  et le temps nécessaire pour en découvrir de nouveaux. Notre objectif
  est d'analyser ces résultats afin de choisir le meilleur protocole que
  nous pourrons utiliser à terme afin de créer un intergiciel
  décentralisé pour les Desktop Grid.}

\MotsCles{Desktop Grid, Évaluation de performance, Découverte de services, ZeroConf, Pastry.}

\section{Introduction}
Les grilles de calcul répondent aux différents besoins de calcul avec un
coût économique. Une variante est la grille de PC (Desktop Grid) où les
n{\oe}uds sont simplement constitués de PC de bureau.  Cette variante
constitue notre cadre de travail. La majorité des intergiciels pour les
grilles de PC actuels sont centralisés. Dans ce contexte, notre travail
consiste à concevoir un intergiciel de Desktop Grid de calcul
décentralisé et basé sur les systèmes P2P.  Pour réaliser cela, nous
voudrions profiter des systèmes P2P décentralisés existants. Dans ce
travail, nous supposons que nous disposons d'un intergiciel de haut
niveau capable à virtualiser le réseau (nous n'avons plus de problèmes
avec les pare-feu et NAT) et nous pouvons exécuter Bonjour, Avahi et
Pastry au dessus de cet intergiciel (Bonjour, Avahi et Pastry
fonctionnent sur un réseau local). Instant Grid~/ Private Virtual
Cluster~\cite{rmnc06} est l'un des candidats pour la virtualisation des
réseaux. Ses principales exigences sont: 1) une configuration simple du
réseau 2) pas de dégradation de la sécurité d'une ressource 3) pas de
besoin de réimplémenter des applications distribuées existantes.  Sous
ces hypothèses, il est raisonnable de s'intéresser aux performances de
Bonjour, Avahi et Pastry.

La découverte de services dans la grille est parmi les principaux défis
à relever. Par exemple, l'intergiciel Globus met en {\oe}uvre le service
publication/découverte, un mécanisme fondé sur le contrôle et la
découverte de Services (MDS-2) ~\cite{fknt02}. Ce protocole utilise un
serveur d'enregistrement centralisé. MDS-2 est en fait une architecture
hiérarchique, mais il est encore très vulnérable avec un seul point de
défaillance. En outre, l'adaptation à la dynamicité de serveurs est un
autre défi pour MDS-2. Une autre alternative consiste à utiliser une
approche décentralisée pour la découverte de services ~\cite{xwy05}.
Récemment, la communauté P2P a élaboré un certain nombre de protocoles
totalement décentralisés, tels que Bonjour ~\cite{book-zeroconf}, Avahi
~\cite{url-avahi} et Pastry ~\cite{rd01,url-freepastry} pour
l'enregistrement, le routage et la découverte dans les réseaux
P2P. L'idée de base derrière ces protocoles est d'établir
l'auto-organisation et la configuration dynamique des organisations
virtuelles lorsque de nouveaux n{\oe}uds rejoignent ces organisations.

L'article est organisé comme suit: dans le paragraphe 2 nous présentons
la notion de grilles de PC et nous illustrons les avantages des systèmes
P2P pour construire de telles grilles. Par la suite, dans le paragraphe
3, nous décrivons la procédure des tests d'expérimentation pour analyser
la performance de Bonjour, Avahi et Pastry. Dans les paragraphes 4, 5 et
6, nous discutons les résultats numériques que nous avons obtenus après
plusieurs expérimentations faites sur la plate-forme Grid'5000 (jusqu'à
308 machines). Une conclusion et des perspectives clôturent l'article.

\section{Les grilles de PC (Desktop Grid)}

Les grilles de calcul (comme Grid'5000) visent à fournir une
infrastructure avec une qualité de service garantie entre des ressources
(relativement) homogènes et des communautés certifiées. En revanche, les
systèmes P2P se concentrent sur la construction d'une très grande
infrastructure entre grandes communautés non certifiées, entre individus
anonymes et entre ressources volatiles. Toutefois, la convergence des
deux systèmes semble naturelle~\cite{fi03}. En fait, les travaux de
recherche sur les systèmes P2P cherchent de plus en plus à fournir des
infrastructures capables de supporter des applications diversifiées
alors que la recherche en grille commence à se soucier de la massification
et de l'extensibilité.

Sous l'hypothèse d'un réseau virtualisé~\cite{rmnc06}, le choix de
Avahi et Bonjour est justifié par le fait qu'ils sont deux
implémentations du protocole ZeroConf (Zero Configuration Networking)
qui a déjà prouvé sa fiabilité dans le domaine des réseaux locaux et
qui peut être extensible sur les réseaux WAN (en utilisant l'envoi
Unicast et le DNS). Parmi les protocoles basés sur une DHT
(Distributed Hash Table) tels que CAN~\cite{retc01} et
CHORD~\cite{setc01}, nous avons choisi Pastry car, d'une part, il
offre les possibilités de réplications et de DHT et, d'autre part, il
existe une implémentation open-source directement exploitable
~\cite{url-freepastry}.

\subsection{Bonjour} Bonjour est une implémentation par Apple du
protocole ZeroConf. Le but est d'obtenir un réseau IP fonctionnel sans
dépendance d'une infrastructure comportant un serveur DHCP et un serveur
DNS ou d'une expertise réseau. Bonjour est architecturé autour de trois
fonctionnalités : il permet l'allocation dynamique d'adresses IP sans
serveur DHCP, il assure la résolution de noms et adresses IP sans
serveur DNS et effectue la recherche des services sans annuaire. Au
niveau technique, Bonjour utilise les adresses de lien-local. Lorsque le
serveur DHCP échoue ou n'est pas disponible, le lien-local permet à un
ordinateur d'obtenir une adresse IP (de type IPv4) tout seul. En IPv4,
l'adresse de lien-local est sélectionnée au moyen d'un générateur
pseudo-aléatoire dans la plage d'adresses de 169.254.1.0 à
169.254.254.255 incluses. La vérification d'unicité d'adresse de
lien-local fonctionne à partir de trois requêtes \emph{ARP probes} qui
sont diffusées sur le lien local. Si l'adresse IP est déjà utilisée (ou
demandée) par une autre machine, alors on tente une autre adresse
fournie par le générateur. Lorsque la machine trouve une adresse libre,
elle diffuse en broadcast deux annonces ARP avec l'adresse source IP
contenant l'adresse sélectionnée. En fait, si à un moment quelconque on
obtient une adresse par DHCP alors on utilise cette adresse et on
abandonne le processus d'auto-configuration sur le lien local. Comme les
adresses de lien-local, lorsque les serveurs DNS ne sont pas disponibles
ou inaccessibles, les machines peuvent encore se référer les uns aux
autres par nom en utilisant le protocole mDNS (Multicast DNS).  Bonjour
utilise le protocole DNS-SD (DNS Service Discovery) pour découvrir les
services publiés dans un réseau local. Puisque DNS-SD est construit au
dessus de DNS (ce sont des requêtes avec le champs TXT qui sont
utilisées), il travaille non seulement avec les mDNS, mais aussi avec
les DNS classiques pour la découverte de services à distance.

\subsection{Avahi}
Avahi est un système qui facilite la découverte de services sur un
réseau local. Il permet à des programmes de publier et de découvrir les
services et les hôtes fonctionnant sur un réseau local sans aucune
configuration spécifique. Avahi est une mise en {\oe}uvre des
spécifications DNS-SD et Multicast DNS de ZeroConf.  Avahi est
essentiellement basé sur l'implémentation Linux de mDNS.  Il utilise
D-Bus (une bibliothèque de communication asynchrone entre processus)
pour la communication entre applications.

\subsection{Pastry} Pastry permet de construire un réseau P2P fondé
sur les clés de hash distribuées, il réalise une correspondance entre
un identifiant et une valeur de hash~\cite{rd01}. Cette correspondance
permet de placer directement une ressource dans le réseau de
pairs. Pastry est une infrastructure générique, évolutive et efficace
pour les applications P2P. Les n{\oe}uds de Pastry forment un réseau de
pairs (ou anneau logique) décentralisé, auto-organisé et tolérant à la
panne. Ceci le différencie des précédents protocoles étudiés. La
construction de l'anneau Pastry commence par la création du n{\oe}ud
\emph{bootstrap}. Par la suite, tous les nouveaux arrivés contactent
le n{\oe}ud bootsrap pour rejoindre l'anneau logique.  Chaque n{\oe}ud a un
identifiant unique (\emph{nodeId}) et une table de routage et un
\emph{leaf set} qui contient les \emph{nodeId} des n{\oe}uds
voisins. Lorsqu'il est présenté avec un message et une clé, un n{\oe}ud
Pastry route efficacement le message au n{\oe}ud qui a le \emph{nodeId}
le plus proche numériquement de la clé, à travers tous les n{\oe}uds
couramment vivants. Pastry est complètement décentralisé, évolutif,
auto-organisé et auto-déterministe; il s'adapte automatiquement à
l'arrivée, le départ ou l'échec de connexion de n{\oe}uds. Pastry tient
compte de la localité dans le réseau; il vise à réduire le temps de
transfert des messages, en minimisant le nombre de sauts de routage
IP.  

\section{Description de la phase expérimentale}
La plate-forme expérimentale utilisée est Grid'5000~\cite{url-grid5000},
hautement reconfigurable et contrôlable, qui rassemble 9 sites
géographiquement distribués en France. Tous les sites sont connectés par
le réseau RENATER (10~Gb/s). Nos tests sont appliqués sur le site
d'orsay, où les n{\oe}uds sont connectées par un réseau de 1~Gb/s. Nous
avons utilisé presque la totalité des machines disponibles dans ce site
(plus de 300 machines). Toutes les machines ont des processeurs AMD
Opterons et des cartes réseaux 1~Gb/s.

Nous représentons les n{\oe}uds par des services pour construire un
réseau virtuelle sur la palteforme Grid'5000. En effet sur chaque
machine nous enregistrons un service, ainsi, si le service
fonctionne alors la machine est connectée sur le reseau, sinon
(nous désactivons ou nous supprimons le service) la machine est
déconnectée.

Notre objectif est d'étudier la capacité de massification et le temps de
réponse des systèmes P2P décrits plus haut. En fait, nous cherchons le
nombre maximal de n{\oe}uds qui peut être supporté par ces outils et le
temps de réponse nécessaire pour découvrir un nouveau n{\oe}ud qui vient
de se connecter sur le réseau (selon l'état de la grille). Les mêmes
critères de mesures sont appliqués pour les trois systèmes.

\subsection{Construction d'un noyau spécifique pour Grid'5000}
Grid'5000 offre une infrastructure avec des noyaux standards. Pour
exécuter nos tests expérimentaux, nous avons personnalisé un noyau
pour supporter Avahi, Bonjour et Free-Pastry. Ainsi, nous avons créé
un noyau spécifique contenant tous les packages nécessaires à
l'exécution de nos codes. Par la suite, en utilisant les deux outils
OAR~\cite{url-oar} et Kadeploy~\cite{url-kadeploy}, nous réservons et
nous déployons le noyau spécifique sur toutes les machines réservées
selon la procédure maintenant classique pour les utilisateurs.

\subsection{Enregistrements séquentiels}
Dans ce test, la première étape est de réserver N n{\oe}uds sur
Grid'5000 (N varie de 100 n{\oe}uds jusqu'à atteindre la valeur de
saturation pour l'enregistrement de services). Le nombre N
représente le maximum de n{\oe}uds qui peut être utilisé pour
l'expérimentation. Chaque n{\oe}ud demande une inscription pour un
service donnée à un instant t. Initialement, tous les n{\oe}uds ont
le code adéquat pour enregistrer un service qui reste inactif tant
que la demande n'est pas encore activée. Soit $\delta$ le temps
d'activation de la demande. Nous activons séquentiellement (chaque
$\delta$ seconds) toutes les demandes (et nous recevons un accusé de
notification). Ainsi, la k\ieme demande sera exécutée à la
$k\times\delta$ unité de temps. Nous augmentons $\delta$ pour
analyser le comportement du système lorsque le délai entre les
événements devient plus grand. Evidemment, au début, le nombre
d'enregistrements est petit, ainsi le temps d'enregistrement est
minime. Nous cherchons à analyser la massification des systèmes sans
saturer le réseau : dans ce type de tests, seulement un envoi
multicast apparaît dans un temps donné.

\subsection{Enregistrements simultanés} 

Dans le premier test, les enregistrements sont exécutés d'une manière
séquentielle, à un instant t chaque n{\oe}ud envoie un seul message, par
conséquent le nombre de communication est limité.  Dans ce type
d'expérimentations, nous «~stressons~» la scalabilté du système
et sa capacité à gérer les communications entre les n{\oe}uds
enregistrés. Par conséquent, nous demandons N (nombre des n{\oe}uds
réservés) enregistrements simultanés et nous mesurons le temps
nécessaire pour terminer la phase d'enregistrement. Si nous obtenos un
temps de réponse raisonnable, nous augmentons la valeur de N jusqu'à
arriver à une valeur de saturation. Autrement, nous cherchons le
nombre maximum de n{\oe}uds qui peut être enregistré lorsque plusieurs
communications (de type multicast) sont présentes sur le réseau.

\subsection{Découverte de  services}

Pour évaluer les trois systèmes candidats, nous avons aussi pris comme
métrique d'évaluation le temps de découverte d'un service. En effet,
le temps de découverte est le temps écoulé entre la fin
d'enregistrement d'un service et sa découverte par le programme
\emph{browser} qui tourne sur une seule machine. Notons que le temps
de réponse dépend du nombre de n{\oe}uds enregistrés.  Le programme
\emph{browser} écoute n'importe quel nouvel événement, i.e. un nouvel
enregistrement ou la suppression d'un service.

Avec les deux types d'enregistrements mentionnés ci-dessus, nous
pouvons analyser l'efficacité de la découverte des services pour
les trois systèmes candidats.

\section{Performances de l'enregistrement des services}

\subsection{Enregistrement des services de Bonjour}

\begin{figure}[ht]\begin{center} 
{\epsfysize=1.5in\epsfbox{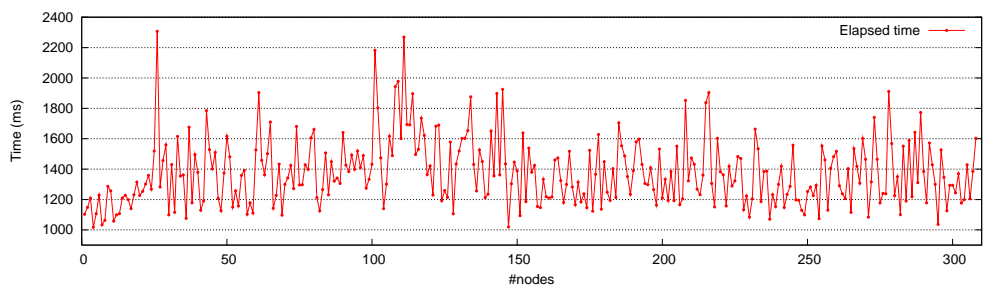}} \caption{Temps écoulé pour
l'enregistrement simultané des services de Bonjour}
\label{simultaneous-registrations-bonjour}
\end{center}\end{figure}

\begin{figure}[ht]\begin{center} 
{\epsfysize=1.5in\epsfbox{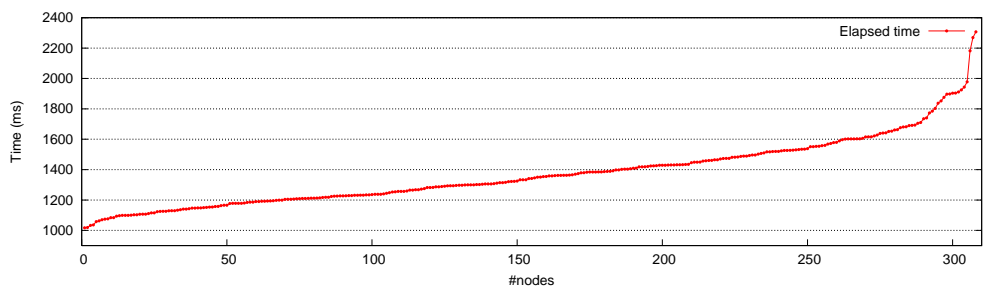}} \caption{Temps écoulé
(triés dans l'ordre croissant) pour l'enregistrement simultané des
services de Bonjour}
\label{simultaneous-registrations-trie-bonjour}
\end{center}\end{figure}

Bonjour commence par faire une requête DHCP pour obtenir une
adresse. En cas d'absence du DHCP, il effectue une (jusqu'à 3)
requête(s) ARP ; or quand on démarre le noyau, une IP est déjà attribuée,
donc on ne peut pas tenir compte du temps écoulé dans cette première
phase. Bonjour fait, ensuite, en sorte d'informer toutes les machines
que le n{\oe}ud a le service en question, il s'agit donc de mettre à jour
les caches ARP. Il pourrait y avoir un coût de gestion supplémentaire
lié à des remplacements dans le cache mais la taille par défaut des
caches ARP est 1040 entrées ce qui est supérieur au nombre de services
que l'on déclenche dans le pire cas de nos expériences. On ne peut pas
raisonnablement incriminer la gestion des caches ARP dans la
dégradation des performances. Bonjour continue par une vérification
que le service est unique. Une première annonce ARP est faite avec une
attente d'une seconde (\emph{Probe wait} = 1s). S'il n'y a pas eu de
réponse dans la seconde (c'est bien le cas, car tous les services ont
été choisis avec un nom unique), alors le nom est considéré comme
unique.

La figure \ref{simultaneous-registrations-bonjour} illustre les temps
d'enregistrement simultané. Jusqu'à 308 machines, le temps écoulé pour
l'enregistrement varie entre 1017~ms et 2307~ms. Sur cette figure
l'axe des abscisses donne le numéro de service et l'axe des ordonnées
sont les temps (chaque machine i enregistre le service i). Il se peut
que le service k s'enregistre avant le service k-p (k et p > 0) ce qui
explique l'aspect «~bruit de fond~» de la
figure~\ref{simultaneous-registrations-bonjour}.

L'enregistrement séquentiel (la figure n'est pas donnée faute de
place), montre une réduction des temps d'enregistrement des
services. En effet, les mesures effectuées sur 308 machines donnent
des temps d'enregistrement compris entre 1015 et 1030~ms (presque
stable). Pour mieux analyser le comportement de l'enregistrement
simultané de services, nous avons tracé la courbe de la figure
~\ref{simultaneous-registrations-trie-bonjour} en triant dans l'ordre
croissant les temps d'enregistrement simultané. En effet, comme nous
l'avons mentionné, dans le test séquentiel le coût varie entre
1015~ms et 1030~ms, dans le cas simultané, on retrouve cette
constante mais il faut en plus accéder au bus à tour de rôle et cela
coûte de l'ordre de 10 à 30~ms, ce qui engendre une augmentation
cumulé du temps d'enregistrement de 1017~ms à 2307~ms.

\subsection{Enregistrement des services de Avahi}

\begin{figure}[ht]\begin{center} 
{\epsfysize=1.5in\epsfbox{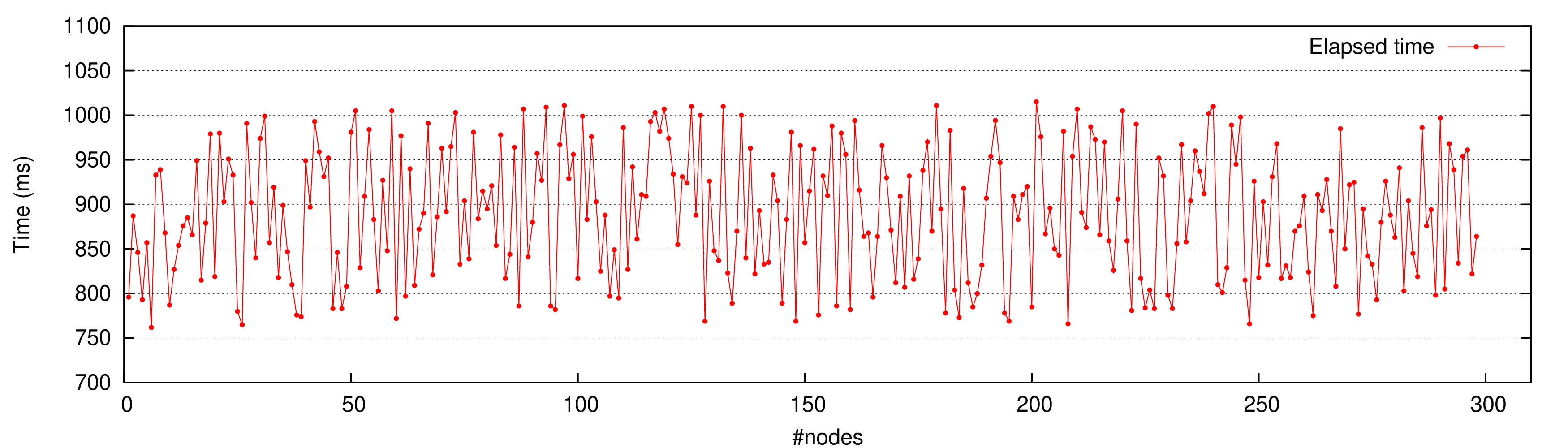}} \caption{Temps écoulé pour
l'enregistrement simultané des services de Avahi}
\label{simultaneous-registrations-avahi}
\end{center}\end{figure}

\begin{figure}[ht]\begin{center} 
{\epsfysize=1.5in\epsfbox{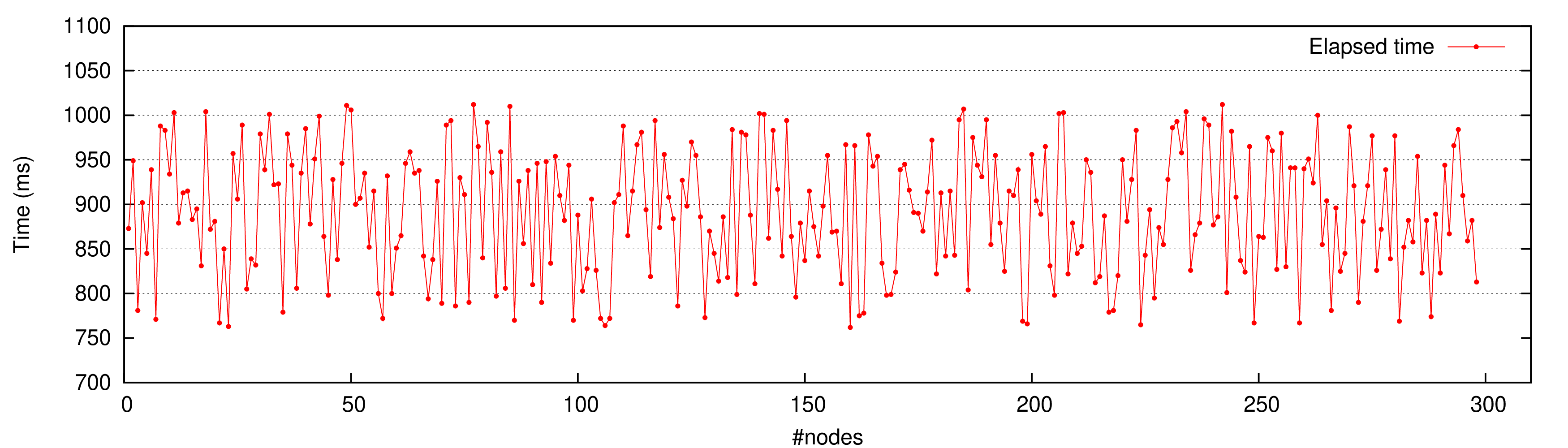}} \caption{Temps écoulé pour
l'enregistrement séquentiel des services de Avahi}
\label{sequential-registrations-avahi}
\end{center}\end{figure}
Sur chaque n{\oe}ud du réseau Avahi, il y'a un démon
\emph{avahi-daemon} qui tourne. Ce démon implémente les deux
protocoles mDNS/DNS-SD de Zeroconf. L'enregistrement d'un service
revient à publier un service avec un envoi multicast à tous les
n{\oe}uds en utlisant D-Bus comme protocole de transport et mettre à
jour les caches de chaque n{\oe}ud. Ces protocoles ont montré une
grande efficacité dans l'enregistrement des services. En effet,
les figures ~\ref{simultaneous-registrations-avahi} et
~\ref{sequential-registrations-avahi} montrent que Avahi donne
presque les mêmes temps d'enregistrements dans les tests
séquentiels et simultanés. Le temps écoulé varie entre 760 et
1110~ms. En comparaison avec Bonjour, Avahi donne de meilleurs
temps d'enregistrements.

\subsection{Enregistrement des services de Pastry}
\begin{figure}[ht]\begin{center} 
{\epsfysize=1.5in\epsfbox{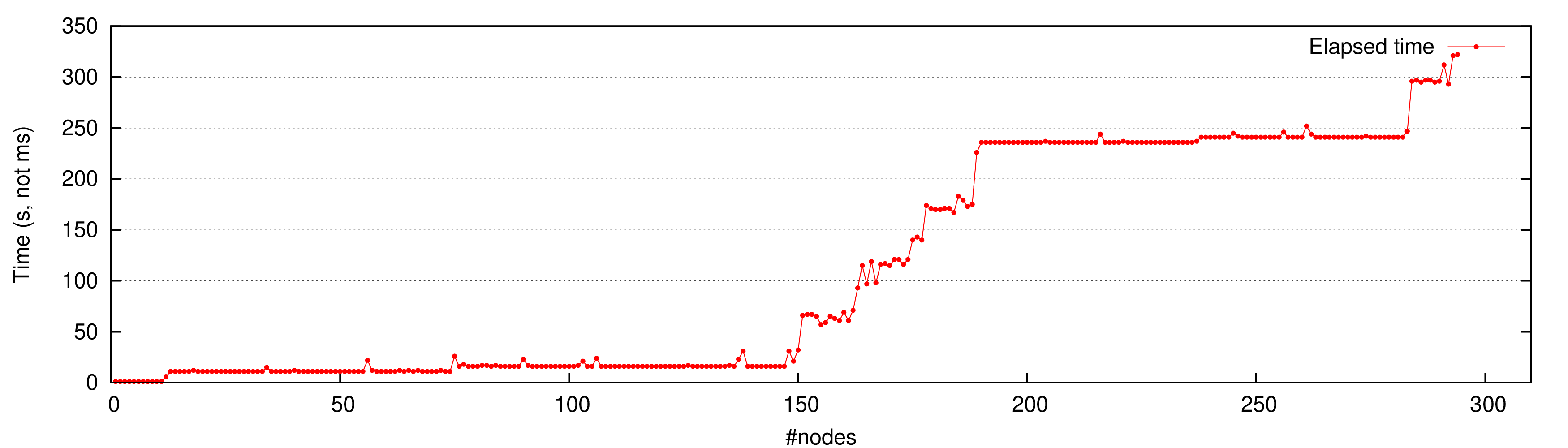}} \caption{Temps écoulé pour
l'enregistrement simultané des services de Pastry}
\label{simultaneous-registrations-pastry}
\end{center}\end{figure}

\begin{figure}[ht]\begin{center}
{\epsfysize=1.5in\epsfbox{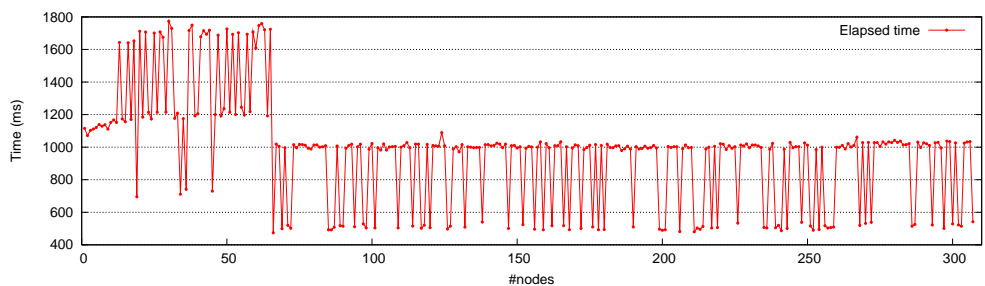}} \caption{Temps écoulé pour
l'enregistrement séquentiel des services de Pastry}
\label{sequential-registrations-pastry}
\end{center}\end{figure}

Contrairement à Avahi et Bonjour, Pastry montre une grande
différence entre les tests d'enregistrements séquentiels et
simultanés. En effet, la figure
\ref{simultaneous-registrations-pastry}, montre que dans
l'enregistrement simultané, jusqu'au 160\ieme  service, le temps
écoulé varie entre 600 et 1000~ms. Au delà, le temps
d'enregistrement augmente d'un enregistrement à un autre pour
atteindre $320\,000$~ms. D'autre part, la
figure~\ref{sequential-registrations-avahi} montre que
l'enregistrement séquentiel donne de meilleurs temps
d'enregistrement. Nous remarquons que Pastry donne des temps
d'enregistrement réduits par rapport à Avahi et Bonjour ( 30\% des
services, parmi 307, sont enregistrés dans l'intervalle de 450 à
550ms). En effet, lorsque nous enregistrons un service, nous
exigeons la connexion à la machine bootstrap (pour ne pas créer
plusieurs anneaux). Ainsi dans l'enregistrement simultané, la
machine bootstrap ne peut pas répondre à toutes les requêtes
simultanées, d'où la croissance très rapide des temps
d'enregistrement depuis le service \no 160. En plus Pastry met à
jour les \emph{leaf sets} pour maintenir la cohérence du système,
et c'est peut être aussi pour cette raison que dans la version
simultanée les temps d'enregistrement arrivent à 320~s. Alors que
dans la version séquentielle, la machine bootstrap ne reçoit
qu'une seule requête chaque minute ; la mise à jour des \emph{leaf
sets} et des tables de routage est ainsi recouvrée d'où des temps
d'enregistrement minimes (2s au maximum).

\section{Performances de la découverte de services}

La deuxième métrique consiste à mesurer le temps nécessaire à la
découverte d'un service enregistré. Ainsi, pour chaque système
(Bonjour, Avahi and Pastry), nous mesurons le temps écoulé entre
la fin de l'enregistrement et l'instant de découverte. Nous
reprenons les mêmes mesures pour les deux types d'enregistrements
simultané et séquentiel. Pour cela, nous consacrons une machine
pour exécuter le programme d'écoute (ou \emph{Browser}) pour la
découverte de services.

\subsection{Comportement de la découverte de Bonjour} Bonjour prouve
une haute performance en découverte de services. En effet, il est
capable de découvrir, dans les deux versions (séquentielles et
simultanées), 307 services enregistrés sur 307 machines (un service
par machine).  Le temps de découverte ne dépasse pas 1 seconde. Cela
nous conduit à affirmer que l'implementation par Apple de DNS-SD
fonctionne très bien et donne des résultats satisfaisants.

\subsection{Comportement de la découverte de Avahi}
\begin{figure}[ht]\begin{center}
{\epsfysize=1.5in\epsfbox{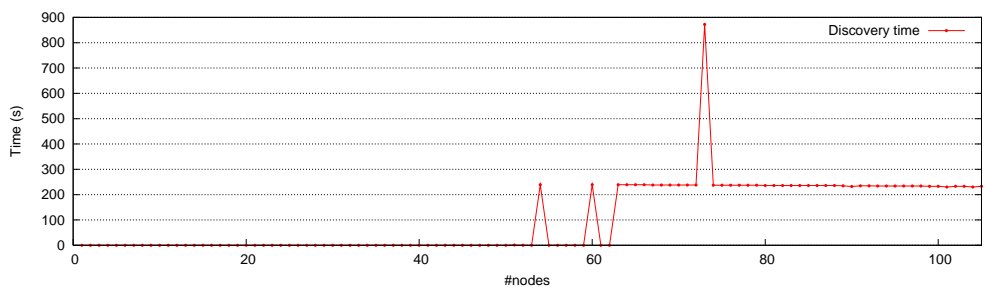}} \caption{Temps de découverte
pour l'enregistrement simultané des services de Avahi}
\label{simultaneous-browsing-avahi}
\end{center}\end{figure}
Comme nous l'avons déjà mentionné, Avahi utilise le démon
\emph{avahi-daemon} et fait une requête via D-Bus pour publier un
service par un paquet multicast. La machine qui lance le programme de
découverte (un programme \emph{Browser} qui reste à l'écoute des
services), perd 60\% des services enregistrés d'une manière
simultanée. En plus, le temps de découverte augmente au delà de 49
enregistrements pour atteindre 900~s dans l'enregistrement du 73\ieme
service. Au delà, le programme de découverte d'Avahi passe autour de
220~s pour découvrir un service enregistré (voir figure
\ref{simultaneous-browsing-avahi}). Contrairement à l'enregistrement
simultané, lorsque nous enregistrons les services d'une manière
séquentielle (chaque minute, nous enregistrons un service), le programme
\emph{Browser} est capable de découvrir plus de services (303 parmi 307
services enregistrés). En plus, le temps de découverte est meilleur
(4~s au maximum) pour 204 services sauf que pour 9 services demandent
des temps de l'intervalle 2835--8686~s.

La machine de découverte (\emph{browser}), qui a le démon
\emph{avahi-daemon} en exécution, n'arrive pas à recevoir, à un
instant donné, tous les envois multicasts émis par les n{\oe}uds qui
ont enregistré les services au même instant (cas simultané), ce
qui peut expliquer la perte des services dans la version
simultanée. Alors que dans la version séquentielle, le
\emph{browser} nécessite plus de temps ($>8686s\simeq{2.5}$
heures) pour la découverte des quatre services perdus.

\subsection{Comportement de la découverte de Pastry}

Dans les deux types d'enregistrements séquentiel et simultané, pastry
donne des temps de réponse rapide de découverte (au plus en 1s). Quant
au nombre de services découverts, pour l'enregistrement simultané, le
\emph{browser} découvre 270 parmi 293 services, alors que pour
l'enregistrement séquentiel, le \emph{browser} découvre 275 des 292
services ce qui correspond à une légère amélioration vis à vis du cas
précédent. Cela peut nous conduis à affirmer que la machine
\emph{Browser} n'arrive pas à récupérer tous les annonces de
publications des services si nous dépassons environ 270
enregistrements.

\section{Synthèse}

La comparaison des trois bibliothèques Bonjour, Avahi et Pastry du point
de vue des temps d'enregistrements simultanés d'environ 300 services
sur la plate-forme Grid'5000 (un service par machine) montre que Avahi
est le meilleur puisqu'il met le moins de temps (le dernier service
enregistré demande 1000 ms). Bonjour nécessite 1300ms de plus pour
enregistrer le dernier service. Pastry donne des temps proches de ceux
mis par Avahi jusqu'à l'enregistrement de 150 services, mais au delà
il met des temps nettement plus grands (jusqu'à 32000ms) que ceux de
Avahi et Bonjour.

Quand nous enregistrons séquentiellement un service sur chaque machine
(nous arrivons jusqu'aux environs de 300 machines), nous pouvons
mentionner qu'il n'y a pas une grande différence entre les trois
bibliothèques. En fait, Bonjour et Avahi donnent des résultats
semblables. Pastry met presque le même temps pour enregistrer 60\% des
services, a besoin de moins de temps pour enregistrer les premiers
30\% mais de plus de temps que Avahi et Bonjour pour les 10\% qui
restent.

\section{Conclusion}

Avec la croissance de la taille des grilles, il est très envisageable
d'utiliser les systèmes P2P reconnus pour leur capacité de massification
et leur gestion de la volatilité. À ce propos, nous avons étudié dans
cet article trois protocoles P2P pour la découverte de ressources qui
sont Bonjour, Avahi et Pastry. Les trois protocoles ont montré de hautes
performances sauf qu'Avahi n'a pas réussi à découvrir tous les services
dans la version simultanée et que Pastry met un temps considérablement
long pour enregistrer tous les services dans la version simultanée. Nous
continuerons à travailler à terme avec les trois protocoles. En effet,
Bonjour est très performant dans l'enregistrement et la découverte dans
les deux versions séquentielle et simultanée.  Pastry a aussi prouvé sa
performance dans l'enregistrement séquentiel et aussi dans la version
simultanée pourvu qu'on ne dépasse pas les 160 services à un instant
donné ce qui ne représente pas vraiment un point de faiblesse si on ne
travaille pas avec 10 millions de n{\oe}uds. En fait, comme nous l'avons
déjà mentionné, étant donné que le grand nombre de connexions à la même
machine bootstrap est à l'origine de ce problème, nous pouvons y
remédier par la création d'un deuxième bootstrap pour initialiser un
deuxième anneau logique. Avahi présente l'avantage d'être libre avec le
code source accessible, ce qui rend son étude beaucoup plus facile que
Bonjour. Du point de vue technique, l'API de ZeroConf n'offre pas toutes
les fonctionnalités pour construire un intergiciel de grille, alors que
Pastry offre une API open source développée avec Java contenant les
fonctions préliminaires nécessaires à l'élaboration de cet
intergiciel. Notre objectif final est de construire un intergiciel de
Desktop Grid basé sur l'un de ces protocoles.

\section*{Remerciements}
Les expériences présentées dans cet article ont été réalisées sur la
plate-forme expérimentale Grid~5000, une initiative du ministère de
l'Enseignement supérieur et de la Recherche à travers l'action
incitative ACI GRID, l'INRIA, le CNRS, RENATER et d'autres partenaires
(see~\cite{url-grid5000}). Nous tenons à remercier Mathieu Jan pour ses
commentaires sur une version préliminaire de cet article.

\shorthandoff{:} \bibliography{biblio}

\begin{thebibliography}{10}

\bibitem{fi03}
Ian Foster and Adriana Iamnitchi.
\newblock On death, taxes, and the convergence of peer-to-peer and grid
  computing.
\newblock In M.~Frans Kaashoek and Ion Stoica, editors, {\em Peer-to-Peer
  Systems II, Second International Workshop, IPTPS 2003}, volume 2735 of {\em
  Lecture Notes in Computer Science}. Springer Verlag, 2003.

\bibitem{fknt02}
Ian Foster, Carl Kesselman, Jeffrey~M. Nick, and Steven Tuecke.
\newblock Grid services for distributed system integration.
\newblock {\em Computer}, 35(6):37--46, 2002.

\bibitem{retc01}
Sylvia Ratnasamy, Paul Francis, Mark Handley, Richard Karp, and Scott Schenker.
\newblock A scalable content-addressable network.
\newblock In {\em SIGCOMM '01: Proceedings of the 2001 conference on
  Applications, technologies, architectures, and protocols for computer
  communications}, pages 161--172, New York, NY, USA, 2001. ACM Press.

\bibitem{rmnc06}
Ala Rezmerita, Tangui Morlier, Vincent N{\'e}ri, and Franck Cappello.
\newblock Private virtual cluster: Infrastructure and protocol for instant
  grids.
\newblock In Wolfgang~E. Nagel, Wolfgang~V. Walter, and Wolfgang Lehner,
  editors, {\em Euro-Par 2006, Parallel Processing, 12th International Euro-Par
  Conference}, volume 4128 of {\em Lecture Notes in Computer Science}, pages
  393--404. Springer, August 2006.

\bibitem{rd01}
Antony I.~T. Rowstron and Peter Druschel.
\newblock Pastry: Scalable, decentralized object location, and routing for
  large-scale peer-to-peer systems.
\newblock In {\em Middleware '01: Proceedings of the IFIP/ACM International
  Conference on Distributed Systems Platforms}, pages 329--350, London, UK,
  2001. Springer-Verlag.

\bibitem{book-zeroconf}
Daniel Steinberg and Stuart Cheshire.
\newblock {\em Zero Configuration Networking: The Definitive Guide}.
\newblock O'Reilly Media, Inc., first edition, December 2005.

\bibitem{setc01}
Ion Stoica, Robert Morris, David Karger, M.~Frans Kaashoek, and Hari
  Balakrishnan.
\newblock Chord: A scalable peer-to-peer lookup service for internet
  applications.
\newblock In {\em SIGCOMM '01: Proceedings of the 2001 conference on
  Applications, technologies, architectures, and protocols for computer
  communications}, pages 149--160, New York, NY, USA, 2001. ACM Press.

\bibitem{xwy05}
Qi~Xia, Weinong Wang, and Ruijun Yang.
\newblock A fully decentralized approach to grid service discovery using
  self-organized overlay networks.
\newblock In Peter M.~A. Sloot, Alfons~G. Hoekstra, Thierry Priol, Alexander
  Reinefeld, and Marian Bubak, editors, {\em Advances in Grid Computing - EGC
  2005, European Grid Conference}, volume 3470 of {\em Lecture Notes in
  Computer Science}, pages 164--172. Springer-Verlag, February 2005.

\bibitem{url-kadeploy}
{KAD}eploy.
\newblock URL: \texttt{http://gforge.inria.fr/projects/kadeploy/}.

\bibitem{url-oar}
{OAR}.
\newblock URL: \texttt{http://gforge.inria.fr/projects/oar/}.

\bibitem{url-avahi}
Avahi.
\newblock URL: \texttt{http://www.avahi.org}.

\bibitem{url-freepastry}
Free{P}astry.
\newblock URL: \texttt{http://www.freepastry.org}.

\bibitem{url-grid5000}
Grid{'}5000.
\newblock URL: \texttt{http://www.grid5000.fr}.

\end{thebibliography}

\end{document}